\documentclass{llncs}
\usepackage{latexsym}
\usepackage{epsf}
\usepackage{theorem}
\usepackage[all]{xy}
\newtheorem{thm}{Theorem}

\theorembodyfont{\parindent=0pt\parskip=2mm\it}
\newtheorem{df}[thm]{Definition}
\theorembodyfont{\rm}

    {\bgroup\parindent=0pt\parskip=2mm plus.4mm\vskip2mm\par\rm
                                           {\rm P~r~o~o~f:}~~}%
    {\hbox{}\hfill$\Box$\par\vskip2mm\egroup}

\def\dldom{\Delta}

\def\id{\mathop{\mathit{Id}}\nolimits}
\def\pf{\mathop{\mathsf{Pf}}\nolimits}

\def\dlbot{\bot}
\def\dlc#1{\mathrm{#1}}
\def\dlall#1#2{\forall #1.#2}

\def\dland#1#2{#1\sqcap#2}

\def\dlnot#1{\neg#1}

\def\dlfd#1#2#3{#1:#2\rightarrow #3}

\def\dlint#1{(#1)^{\cal I}}
\def\dls{\sqsubseteq}

\def\dld{\mbox{$\cal DLFD$}}     % functions, booleans, and pfds
\def\mf{\models_{\mbox{\rm\scriptsize fin}}}

\begin{document}
\pagestyle{plain}
%\abovedisplayskip=4pt
%\belowdisplayskip=4pt

%% lncs bug
%\let\xxx=\subsubsection
%\def\subsubsection#1{\xxx{#1.}}
\let\xxx=\paragraph
\def\paragraph#1{\xxx{#1.}}

\title{Undecidability of Finite Model\\ Reasoning in $\dld$}
\author{David Toman~~~~and~~~~Grant Weddell}
\institute{
  Cheriton School of Computer Science\\
  University of Waterloo, Canada\\
           \tt$\{$david,gweddell$\}$@uwaterloo.ca}
\date{}
\maketitle

\begin{abstract}
  We resolve an open problem concerning finite logical implication for
  path functional dependencies (PFDs). 
  This note is an addendum to \cite{TW05a}.
\end{abstract}

In this note we show that the finite logical implication for
description logics endowed with PFDs is undecidable. This result complements
the decidability of the unrestricted problem that is complete for EXPTIME
\cite{TW05b,TW01}.

\section{Preliminaries}\label{s:def}

%\begin{df}[Syntax and Semantics of \dld]\label{d:dl}~\\
\begin{df}[Description Logic \dld]\label{d:dl}
  Let $F$ and $C$ be sets of feature names and primitive concept
  names, respectively. A {\em path expression\/} is defined by the
  grammar ``\,$\pf ::= f.\pf | \id$'' for $f\in F$.  We define derived
  {\em concept descriptions\/} by a second grammar on the left-hand-side
  of Figure~\ref{fig:DL}.  A concept description obtained by using the
  final production of this grammar is called a {\em path-functional
  dependency} (PFD).
  
  An {\em inclusion dependency\/} $\cal C$ is an expression of the
  form $D\dls E$.
  A {\em terminology} $\cal T$ consists of a finite set of inclusion
  dependencies.

%\begin{figure}[t]
%\hrule
%\
%\begin{center}\small%\footnotesize
%\tabcolsep=3pt\begin{tabular}{rcl@{~~~~~}l}
%\multicolumn{3}{c}{{\sc Syntax}}&
%  \multicolumn{1}{c}{{\sc Semantics: Defn of ``$\dlint{\cdot}$''}}\\[1mm]
%$D$ & $::=$ & $C$ 
%        & $\dlint{C}\subseteq\dldom$ \\
%& $|$ & $\dland{D_1}{D_2}$  
%        & $\dlint{D_1}\cap\dlint{D_2}$ \\
%%& $|$ & $\dlor{D_1}{D_2}$
%%        & $\dlint{D_1}\cup\dlint{D_2}$ \\
%& $|$ & $\dlnot{D}$ 
%        & $\dldom \setminus \dlint{D}$ \\
%& $|$ & $\dlall{f}{D}$ 
%        & $\{x : \dlint{f}(x) \in \dlint{D}\}$ \\
%& $|$ & $\dlat{f}{D}$ 
%        & $\{\dlint{f}(x): x \in \dlint{D}\}$ \\[2mm]
%$E$ & $::=$ & $D$ \\
%%& $|$ & $\dland{E_1}{E_2}$
%%        & $\dlint{E_1}\cap\dlint{E_2}$ \\
%%& $|$ & $\dlall{f}{E}$ 
%%        & $\{x : \dlint{f}(x) \in \dlint{E}\}$ \\
%% & $|$ & $\dleq{\pf_1}{\pf_2}$
%%         & $\{x : \dlint{\pf_1}(x)=\dlint{\pf_2}(x)\}$ \\
%& $|$ & $\dlfd{D}{\pf_{1}, ... , \pf_{k}}{\pf}$
%        & $\{x : \forall\, y \in \dlint{D}.$ \\
%&&& $\;~~\bigwedge_{i=1}^k \dlint{\pf_i}(x)=\dlint{\pf_i}(y)
%                     \Rightarrow \dlint{\pf}(x)=\dlint{\pf}(y)\}$\\
%\end{tabular}
%\end{center}
%\hrule
%\caption{\sc Syntax and Semantics of \dld.}\label{fig:DL}
%\end{figure}
\begin{figure}[ht]
\hrule
\begin{center}%\small%\footnotesize
\tabcolsep=3pt\begin{tabular}{rcl@{~~~}l}
~\\[-2mm]
\multicolumn{3}{c}{{\sc Syntax}}&
  \multicolumn{1}{c}{{\sc Semantics: ``$\dlint{\cdot}$''}}\\[1mm]
$D$ & $::=$ & $C$ 
        & $\dlint{C}\subseteq\dldom$ \\
& $|$ & $\dland{D_1}{D_2}$  
        & $\dlint{D_1}\cap\dlint{D_2}$ \\
%& $|$ & $\dlor{D_1}{D_2}$
%        & $\dlint{D_1}\cup\dlint{D_2}$ \\
& $|$ & $\dlnot{D}$ 
        & $\dldom \setminus \dlint{D}$ \\
& $|$ & $\dlall{f}{D}$ 
        & $\{x : \dlint{f}(x) \in \dlint{D}\}$ \\[1mm]
%& $|$ & $\dlat{f}{D}$ 
%        & $\{\dlint{f}(x): x \in \dlint{D}\}$ \\
~\\
$E$ & $::=$ & $D$ \\
& $|$ & $\dland{E_1}{E_2}$
        & $\dlint{E_1}\cap\dlint{E_2}$ \\
%& $|$ & $\dlall{f}{E}$ 
%        & $\{x : \dlint{f}(x) \in \dlint{E}\}$ \\
% & $|$ & $\dleq{\pf_1}{\pf_2}$
%         & $\{x : \dlint{\pf_1}(x)=\dlint{\pf_2}(x)\}$ \\
& $|$ & $\dlfd{D}{\pf_{1}, ... , \pf_{k}}{\pf}$
        & $\{x : \forall\, y \in \dlint{D}.\bigwedge_{i=1}^k  $\\
&&& $\;~~\dlint{\pf_i}(x)=\dlint{\pf_i}(y)
                      \rightarrow \dlint{\pf}(x)=\dlint{\pf}(y)\}$\\
\end{tabular}
\end{center}
\hrule
\caption{\sc Syntax and Semantics of \dld.}\label{fig:DL}
\end{figure}
The {\em semantics\/} of expressions is defined with respect to a
structure $(\dldom, \cdot^{\cal I})$, where $\dldom$ is a
domain of ``objects'' and $\dlint{.}$ an interpretation function that
fixes the interpretations of primitive concepts $C$ to be subsets of
$\dldom$ and primitive features $f$ to be total functions
$\dlint{f}:\dldom\rightarrow\dldom$.  The interpretation is extended
to path expressions, $\dlint{\id} = \lambda x.x$, $\dlint{f.\pf} =
\dlint{\pf}\circ\dlint{f}$ and derived concept descriptions $D$ and
$E$ as defined on the right-hand-side of Figure~\ref{fig:DL}.

An interpretation {\em satisfies an inclusion dependency $D\dls E$\/}
if $\dlint{D}\subseteq\dlint{E}$. The {\em logical implication
  problem\/} asks if ${\cal T} \models D\dls E$ holds; that is, if
$\dlint{D}\subseteq\dlint{E}$ for all interpretations that satisfy all
constraints in $\cal T$.
\end{df}
%
%We allow ourselves some liberties with more abbreviated notation in
%the remainder of the paper. In particular, trailing $\id$'s in
%path expressions will be abstracted, and the adjacent placement of
%two or more such expressions, $\pf_1 \pf_2 \ldots$, denotes
%a path expression corresponding to their composition. For example,
%we assume $\pf_1 \pf_2 g \pf_3$ denotes the path expression
%$f.g.h.\id$ when the $\pf_i$ have the respective forms $f.\id$, $\id$
%and $h.\id$.  Finally, we write $\dlat{\pf_1 \pf_2}{D}$ as shorthand
%for $\dlat{\pf_2}{\dlat{\pf_1}{D}}$.
%
%For the remainder of the paper, we use the following abbreviated notation: 
%$\dlall{\pf}{\dlc{D}}$ is shorthand for
%$\dlall{f_1}{\dlall{f_2}{\ldots\dlall{f_k}\dlc{D}}}$, and
%$\dlat{\pf}{\dlc{D}}$ for
%$\dlat{f_k}{\ldots\dlat{f_2}{\dlat{f_1}{\dlc{D}}}}$ for
%$\pf=f_1.f_2.\cdots .f_k.\id$. We also identify single feature names $f$
%with path descriptions $f.\id$ and allow concatenation of path
%descriptions, $\pf\pf'$ to denote their composition $\pf\circ\pf'$.

In addition, we classify constraints by the description on their
right-hand side as \emph{PFDs}, when the right-hand side is of the form
$\dlfd{D}{\pf_1,\ldots,\pf_k}{\pf}$, and as \emph{simple constraints}
otherwise.

\section{Undecidability}\label{s:undec}
We show a reduction of a tiling problem to the finite logical implication
problem for \dld\ using a construction similar to that presented in
\cite{TW05a}. In this earlier work, the \emph{unrestricted} tiling
problem that asks if an infinite tiling exists was used. In our case,
we rely on a finite version of a similar problem that remains
undecidable.

A tiling problem $U$ is a triple $(T,H,V)$ in which $T$ is a finite
set of tile types and $H,V\subseteq T\times T$ are a pair of binary
relations.  A solution to a tiling problem is an assignment of tiles
to a two-dimensional surface that satisfies the $H$ and $V$
relations.\footnote{The types of tiles placed side-by-side or one
  above the other must appear in $H$ and in $V$, respectively.}  For
example, a solution for an \emph{unrestricted upper quadrant tiling
  problem} is a function $t:\textbf{N}\times\textbf{N}\rightarrow T$
such that $(t(i,j),t(i+1,j))\in H$ and $(t(i,j),t(i,j+1))\in V$ for
all $i\in\textbf{N}$. This problem can simulate a \emph{Turing machine
  looping} problem, which is not decidable \cite{Ber66,Bo97}.
Similarly, determining if a finite $n\times m$ tiling exists, for some
$n,m>0$, (given an initial tile placed in a lower left corner) implies
that a Turing machine halts (starting from an empty tape).  To reduce
notation in the following, we consider more particularly the tiling
problem of a \emph{finite torus}. (It is straightforward but tedious
to show how to simulate an $n\times m$ tiling with a finite torus
tiling.)

The main step in the reduction is to establish an integer
torus in which an arbitrarily large finite rectangle can be
embedded.  This can be achieved, e.g.,  as follows.
\begin{figure*}[t]
  $$\xymatrix@R=5mm@C=5mm@H=4mm@W=4mm{
&&&&&&&&&&\\
& *=<6mm>[o][F-]{} &
& *=<6mm>[o][F-]{B}\ar[dd]^{f}\ar[ll]_{h}\ar[rr]^{i} 
       \ar@/_/@{<-}[ul]^{b'}\ar@/_/@{<-}[dr]^{b} &
& *=<6mm>[o][F-]{} &
& *=<6mm>[o][F-]{D}\ar[dd]^{g}\ar[ll]_{i}\ar[rr]^{h} 
       \ar@/_/@{<-}[ul]^{d}\ar@/_/@{<-}[dr]^{d'} &
& *=<6mm>[o][F-]{} & \\
&&&&&&&&&&\\
& *=<6mm>[o][F-]{A}\ar[rr]^{f}\ar[uu]^{h}\ar[dd]_{i}
       \ar@/_/@{<-}[ul]^{a'}\ar@/_/@{<-}[dr]^{a} &
& *=<6mm>[o][F-]{X}
       \ar@/_/[dl]_{a}\ar@/_/[dr]_{b}\ar@/_/[ur]_{c}\ar@/_/[ul]_{d} &
& *=<6mm>[o][F-]{C}\ar[ll]^{f}\ar[rr]^{g}\ar[uu]^{i}\ar[dd]^{h}
       \ar@/_/@{<-}[ul]^{c}\ar@/_/@{<-}[dr]^{c'}  &
& *=<6mm>[o][F-]{Y} 
       \ar@/_/[ul]_{a'}\ar@/_/[ur]_{b'}\ar@/_/[dl]_{c'}\ar@/_/[dr]_{d'} &
& *=<6mm>[o][F-]{A}\ar[ll]^{g}\ar[uu]^{h}\ar[dd]_{i} 
       \ar@/_/@{<-}[ul]^{a'}\ar@/_/@{<-}[dr]^{a} &
\\
&&&&&&&&&&\\
& *=<6mm>[o][F-]{} &
& *=<6mm>[o][F-]{D}\ar[uu]^{f}\ar[ll]^{i}\ar[rr]_{h} 
       \ar@/_/@{<-}[ul]^{d}\ar@/_/@{<-}[dr]^{d'} &
& *=<6mm>[o][F-]{} &
& *=<6mm>[o][F-]{B}\ar[uu]^{g}\ar[ll]^{h}\ar[rr]_{i}  
       \ar@/_/@{<-}[ul]^{b'}\ar@/_/@{<-}[dr]^{b} &
& *=<6mm>[o][F-]{} & \\
&&&&&&&&&&\\
}$$
\caption{\sc Interpretation Forming a Tiled Torus.}\label{f:tile}
\end{figure*}
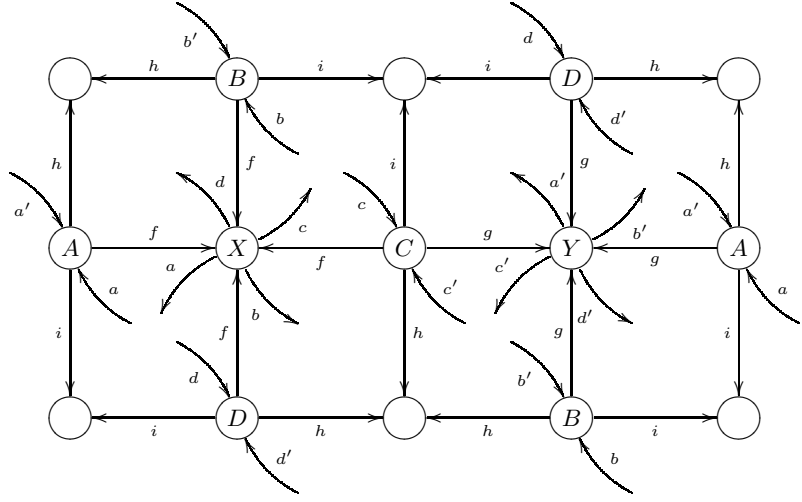
\begin{enumerate}\itemsep=1pt
\item Introduce four disjoint concepts, $A$, $B$, $C$ and $D$, to denote
  cell edges.
% A&B < 0  A&C < 0  A&D < 0  ... C&D < 0 (A,B,C,D paiwise disjoint )
\begin{eqnarray*}%\small
  \dland{\dlc{A}}{\dlc{B}}\dls\dlbot,~~
  \dland{\dlc{A}}{\dlc{C}}\dls\dlbot,~~ \ldots,~~
  \dland{\dlc{C}}{\dlc{D}}\dls\dlbot
\end{eqnarray*}
\item Map grid cells to concepts $\dlc{X}$ and $\dlc{Y}$ that have four
incoming $f$ and $g$ attributes, respectively.
% X < A@{f} & B@{f} & C@{f} & D@{f}
% Y < A@{g} & B@{g} & C@{g} & D@{g}      (every X,Y have 4 neighbors)
\begin{eqnarray*}%\small
&&
\dlc{X}\dls\dland{\dland{\dlall{a}{\dlc{A}}}{\dlall{b}{\dlc{B}}}}
                 {\dland{\dlall{c}{\dlc{C}}}{\dlall{d}{\dlc{D}}}},~~~~~
\dlc{Y}\dls\dland{\dland{\dlall{a'}{\dlc{A}}}{\dlall{b'}{\dlc{B}}}}
                 {\dland{\dlall{c'}{\dlc{C}}}{\dlall{d'}{\dlc{D}}}}\\[2mm]
% &&\dlc{X}\dls\dlfd{X}{a.f}{\id},~
%   \dlc{X}\dls\dlfd{X}{b.f}{\id},~
%   \dlc{X}\dls\dlfd{X}{c.f}{\id},~
%   \dlc{X}\dls\dlfd{X}{d.f}{\id}\\
% &&\dlc{Y}\dls\dlfd{Y}{a'.g}{\id},~
%   \dlc{Y}\dls\dlfd{Y}{b'.g}{\id},~
%   \dlc{Y}\dls\dlfd{Y}{c'.g}{\id},~
%   \dlc{Y}\dls\dlfd{Y}{d'.g}{\id}
&&\dlc{X}\dls\dlfd{\dlc{X}}{a}{\id},~
  \dlc{X}\dls\dlfd{\dlc{X}}{b}{\id},~
  \dlc{X}\dls\dlfd{\dlc{X}}{c}{\id},~
  \dlc{X}\dls\dlfd{\dlc{X}}{d}{\id}\\
&&\dlc{Y}\dls\dlfd{\dlc{Y}}{a'}{\id},~
  \dlc{Y}\dls\dlfd{\dlc{Y}}{b'}{\id},~
  \dlc{Y}\dls\dlfd{\dlc{Y}}{c'}{\id},~
  \dlc{Y}\dls\dlfd{\dlc{Y}}{d'}{\id}\\[3mm]
&&\dlc{A}\dls\dland{\dlall{f}{\dlc{X}}}{\dlall{g}{\dlc{Y}}},~
  \dlc{B}\dls\dland{\dlall{f}{\dlc{X}}}{\dlall{g}{\dlc{Y}}},~
  \dlc{C}\dls\dland{\dlall{f}{\dlc{X}}}{\dlall{g}{\dlc{Y}}},~
  \dlc{D}\dls\dland{\dlall{f}{\dlc{X}}}{\dlall{g}{\dlc{Y}}}\\[2mm]
&&\dlc{A}\dls\dlfd{\dlc{A}}{f}{\id},~
  \dlc{B}\dls\dlfd{\dlc{B}}{f}{\id},~
  \dlc{C}\dls\dlfd{\dlc{C}}{f}{\id},~
  \dlc{D}\dls\dlfd{\dlc{D}}{f}{\id}\\
&&\dlc{A}\dls\dlfd{\dlc{A}}{g}{\id},~
  \dlc{B}\dls\dlfd{\dlc{B}}{g}{\id},~
  \dlc{C}\dls\dlfd{\dlc{C}}{g}{\id},~
  \dlc{D}\dls\dlfd{\dlc{D}}{g}{\id}
\end{eqnarray*}
\item Ensure that squares are formed by adding the following%
\footnote{Note that the asymmetric PFDs can be simulated by the following
$\dlc{A}\dls\dlc{AB}$, $\dlc{B}\dls\dlc{AB}$, and
$\dlc{AB}\dls\dlfd{\dlc{AB}}{f}{h}$ (and similarly for the remaining
cases).}.
% A < B:{f}->{h}     A < B:{g}->{i}
% B < C:{f}->{i}     B < C:{g}->{h}
% C < D:{f}->{h}     C < D:{g}->{i}
% D < A:{f}->{i}     D < A:{g}->{h}      (enforces "squares")
\begin{eqnarray*}%\small
&&\dlc{A}\dls\dlfd{\dlc{B}}{f}{h},~~
\dlc{B}\dls\dlfd{\dlc{C}}{f}{i},~~
\dlc{C}\dls\dlfd{\dlc{D}}{f}{h},~~
\dlc{D}\dls\dlfd{\dlc{A}}{f}{i},\\
&&\dlc{A}\dls\dlfd{\dlc{B}}{h}{f},~~
\dlc{B}\dls\dlfd{\dlc{C}}{i}{f},~~
\dlc{C}\dls\dlfd{\dlc{D}}{h}{f},~~
\dlc{D}\dls\dlfd{\dlc{A}}{i}{f},\\
&&\dlc{A}\dls\dlfd{\dlc{B}}{g}{i}, ~~
\dlc{B}\dls\dlfd{\dlc{C}}{g}{h}, ~~
\dlc{C}\dls\dlfd{\dlc{D}}{g}{i}, ~~
\dlc{D}\dls\dlfd{\dlc{A}}{g}{h},\\
&&\dlc{A}\dls\dlfd{\dlc{B}}{i}{g}, ~~
\dlc{B}\dls\dlfd{\dlc{C}}{h}{g}, ~~
\dlc{C}\dls\dlfd{\dlc{D}}{i}{g}, ~~
\dlc{D}\dls\dlfd{\dlc{A}}{h}{g}
\end{eqnarray*}
The dependencies
$\dlc{X}\dls\dland{\dlall{a}{\dlc{A}}}{\dlfd{\dlc{X}}{a}{\id}}$ and
  $\dlc{A}\dls\dland{\dlall{f}{\dlc{X}}}{\dlfd{\dlc{A}}{a}{\id}}$
    induce, as a finite logical consequence, an incoming $f$ feature
    originating in an $\dlc{A}$ object for \emph{every} $\dlc{X}$
    object. The same holds for $\dlc{B}$, $\dlc{C}$, and $\dlc{D}$
    objects; hence every $\dlc{X}$ object has four incoming $f$
    features.
\item And force squares to extend to the right and up by including
  the following.
\begin{eqnarray*}%\small
  \dlc{A}\dls\dlall{g}{\dlc{Y}},~~~ 
  \dlc{B}\dls\dlall{g}{\dlc{Y}},~~~ 
  \dlc{C}\dls\dlall{f}{\dlc{X}},~~~
  \dlc{D}\dls\dlall{f}{\dlc{X}}
\end{eqnarray*}
\end{enumerate}
The accumulated effect of these inclusion dependencies on an
interpretation is illustrated in Figure~\ref{f:tile}.

The adjacency rules for the instance $U$ of the tiling
problem can now be captured as follows:
\begin{eqnarray*}%\small
&&\dland{\dlc{A}}{\dlall{g}{T_i}}\dls
        \dlall{f}{\textstyle\bigsqcup_{(t_i,t_j)\in V} T_j}, ~~~~
  \dland{\dlc{C}}{\dlall{f}{T_i}}\dls
        \dlall{g}{\textstyle\bigsqcup_{(t_i,t_j)\in V} T_j}\\
&&\dland{\dlc{B}}{\dlall{f}{T_i}}\dls
        \dlall{g}{\textstyle\bigsqcup_{(t_i,t_j)\in H} T_j}, ~~~~
  \dland{\dlc{D}}{\dlall{g}{T_i}}\dls
        \dlall{f}{\textstyle\bigsqcup_{(t_i,t_j)\in H} T_j},
\end{eqnarray*}
where $T_i$ corresponds to a tile type $t_i\in T$; we assume
$\dland{T_i}{T_j}\dls\dlbot$ for all $i<j$.

The combination of all the above comprise a terminology ${\cal T}_U$
associated with a tiling problem $U$.
Now, $U$ admits a finite solution iff 
$${\cal T}_U\not\mf \dland{\dlc{X}}{T_0} \dls\dlbot,$$
where $T_0$ is an initial tile. And since the halting problem can be
reduced to the existence of a finite tiling, we therefore have the
following.
\begin{thm}
The finite logical implication problem for $\dld$ is undecidable.
\end{thm}
Consequently, finite satisfiability $\dld$ knowledge bases is also undecidable.
Note that the construction uses only unary keys (functionality) and
unary functional dependencies and does not need the full power of PFDs.

%Hence the result also shows that extending the decision procedure for unary
%inclusion dependencies in presence of functional dependencies
%\cite{CKV90} with Boolean operations (e.g., with respect to primary
%keys) is not possible.
%small
\bibliography{bib}
%\bibliography{icdt05}
\bibliographystyle{plain}

\end{document}